\documentclass[onecolumn,usenatbib]{mn2e}

\usepackage{graphicx}
\usepackage{natbib}
\begin{document}

\title{Modelling the spin equilibrium of neutron stars in LMXBs without
gravitational radiation} 

\author[N. Andersson et al]
{N. Andersson$^1$, K. Glampedakis$^1$, B. Haskell$^1$ and A.L.Watts$^2$ \\
$^1$ School of Mathematics, University of Southampton, 
Southampton, SO17 1BJ, United Kingdom \\ $^2$ Laboratory for High
Energy Astrophysics, NASA Goddard Space Flight Center, Greenbelt, MD
20771, USA}

\maketitle

\begin{abstract}
In this paper
we discuss the spin-equilibrium of accreting neutron stars in LMXBs. 
We demonstrate that, when combined with a naive spin-up torque, 
the observed data leads to inferred magnetic 
fields which are at variance with those of galactic millisecond radiopulsars.
This indicates the need for either additional spin-down torques (eg. gravitational radiation)
or an improved accretion model. We show that a simple
consistent accretion model can be arrived at by 
accounting for radiation pressure in rapidly accreting systems (above a few percent of the Eddington accretion rate). 
In our model the inner disk region is thick and significantly sub-Keplerian, 
and the estimated equilibrium periods are such that the LMXB neutron stars
have properties that accord well with the galactic millisecond radiopulsar sample. 
The implications for future gravitational-wave observations are also discussed briefly.
\end{abstract}

\maketitle

\section{Introduction}

In the last few years the evidence in favour of the 
notion that neutron stars are spun up to millisecond periods
in accreting systems has strengthened significantly. 
The discovery of the millisecond X-ray pulsar SAX J1808.4-3658 in 
an accreting 
low-mass X-ray binary (LMXB) provided the long anticipated missing link 
between the general LMXB population and the  millisecond radio pulsars.
Since then four similar system has been observed, further strengthening
the connection \citep{wijnands03}. Furthermore, the 
link between the twin-peak separation of the kHz quasiperiodic oscillations
(QPOs) 
seen in a number of systems and the spin of the neutron star has become 
somewhat clearer (although the underlying mechanism is still under debate) 
following the observation of QPOs in SAX J1808.4-3658. It now appears as if
the QPO separation is either equal to  or half the spin period \citep{Miller}.

When the first indications of rapidly spinning neutron stars in LMXBs
were discussed more than five years ago, the results suggested 
that the systems were clustered in a surprisingly narrow range of spin 
frequencies 250-370~Hz. As such spin rates are far below the 
predicted break-up limit of about 1~kHz, the data  pointed
towards the presence of a mechanism that could counteract the 
accretion spin-up torque.   
The obvious candidate --- the interaction between the accretion disk
and the magnetosphere of the neutron star --- was discussed by 
\citet{wz97}. Their results seemed to indicate the need for an
unanticipated link between the accretion rate and the magnetic 
field strength. Since there is no reason to expect such fine-tuning in 
these systems, \citet{bildsten98} argued that an additional 
spin-down mechanism may be in operation. He proposed that this 
torque could be provided by gravitational-wave emission, and 
that the required asymmetries would be induced in the neutron star 
crust by accretion. This idea echoed earlier suggestions 
by \citet{pap} and \citet{wag} that neutron 
stars may reach a spin-equilibrium with gravitational waves balancing the 
accretion torque. 

The possibility that accreting neutron stars may radiate gravitational waves
is of great interest given the generation of groundbased interferometers 
(LIGO, GEO600, TAMA300 and VIRGO) that is now reaching design sensitivity.
It has been recognized that there are three distinct mechanisms that may 
be able to generate gravitational waves at the required rate. 
First of all, a more detailed study 
by \citet{ush} suggests that the 
accretion induced crustal asymmetry proposed by Bildsten remains 
viable. The second possibility is that the stars spin fast 
enough that the gravitational-wave driven instability of the r-mode
oscillations in the neutron star core is activated (see \citet{narev} for references). 
Finally, \citet{cutler} has suggested that an internal toroidal magnetic field 
could lead to unstable free precession resulting in the star
``tipping over'' and becoming an orthogonal rotator, an 
efficient gravitational-wave source. 

The present investigation is motivated by the following facts:
\begin{itemize}
\item The observational data has improved considerably since the original 
discussions in 1997-98.  We now know that the LMXBs are not clustered
in as narrow a range of spins as was originally thought, the 
current range being 250-620~Hz. It is relevant to ask to what extent the
more recent data supports the need for an additional spin-down torque, eg.
gravitational radiation, in these systems.  

\item A question that does not seem to have attracted much interest
concerns whether a more refined model of the interaction between the accretion
disk and the magnetosphere of the neutron star would be able to provide a satisfactory  description  of the LMXBs. 
After all, many important physical mechanisms 
were not accounted for in the analysis of \citet{wz97} and it may be wise not to refine the 
various gravitational-wave scenarios before their relevance
is investigated.

\item If we suppose that the LMXBs radiate gravitational waves at a 
significant level, then we need to address many difficult issues associated 
with the detection of such signals. A key issue concerns the spin-evolution of 
the system. Can we assume that the spin-period remains stable on a time-scale
of a few months? After all, the signal needs to be integrated for at least two weeks in order to be detectable in the noisy data-stream. If the system 
tends to wander, as the data for slower spinning systems suggests 
\citep{bildsten97}, then we need to be able to model the accretion torque 
reliably.    
\end{itemize}

In this paper we aim to address the second of these points. We discuss the 
argument that an additional spin-down torque is needed in the LMXBs, and 
provide a more detailed accretion model that is able to describe 
these systems without particular fine-tuning of the magnetic field.
From this exercise we conclude that it may not be appropriate to assume 
that the neutron stars in LMXBs radiate gravitational waves at a rate that 
exactly balances the accretion spin-up torque expected for a non-magnetic star. 
We do not think this should 
be taken as meaning that these systems are irrelevant for
gravitational-wave physics. The proposed mechanisms for generating 
gravitational radiation should certainly still work. Yet, our discussion 
makes it clear that modelling these systems is significantly more difficult
than has been assumed so far (at least in the gravitational-wave community). 
Of course, by constructing a more detailed 
accretion model, we are beginning to address this issue. 

\section{LMXBs and the ``standard'' accretion model} 

In the simplest models of accreting non-magnetic stars it is assumed that 
matter falling onto the surface of the star provides a torque proportional 
to the angular momentum associated with a Keplerian orbit at the stars equator;
\begin{equation}
N \approx  \dot{M} \sqrt{GMR}
\label{nonmagtorque}\end{equation}
where $M$ is the mass, $R$ the radius and $\dot{M}$ the accretion rate. 
Despite it being well-known that this torque only provides an 
order-of-magnitude estimate, it has been used in most studies of 
gravitational waves from LMXBs so far. 
The line of reasoning has been 
that, if the neutron star is at spin equilibrium, then  the radiated
gravitational waves provide an equal and opposite torque. The strength 
of the gravitational waves can be inferred from the X-ray luminosity,
since  (assuming that the gravitational potential 
released by the infalling matter 
is radiated as X-rays)
\begin{equation}
L_X \approx { GM \dot{M} \over R} 
\end{equation} 
provides a link between the observations and the mass accretion rate. 

Accretion onto a magnetised star is different since the pressure 
of the infalling gas is counteracted by the magnetic pressure. By
balancing these two pressures (for spherical infall) one obtains the
so-called magnetosphere radius
\begin{equation}
R_M \approx 7.8 \left( {B_0 \over 10^{8} \mbox{ G} } \right)^{4/7} 
\left( {R \over 10 \mbox{ km} } \right)^{12/7} \left( {M \over 1.4 M_\odot } \right)^{-1/7}
\left( {\dot{M} \over \dot{M}_\mathrm{Edd}  } \right)^{-2/7}  \mbox{ km}
\label{magneto}\end{equation}
inside which with the flow of matter is likely to be dominated by the magnetic field.  
For a strongly magnetised star the magnetic field is expected to channel the accreting matter 
onto the polar caps \citep{frank}, while the situation may be more complex for a weakly 
magnetised object. 

An approximation of the maximum accretion rate we should expect follows
from balancing the pressure due to spherically infalling gas to that
of the emerging radiation. This leads to the Eddington limit;
\begin{equation}
\dot{M}_\mathrm{Edd} 
\approx 1.5 \times 10^{-8} \left( { R \over 10\mbox{ km}} \right) { M_\odot \over \mbox{yr} }
\end{equation}
with associated X-ray luminosity
\begin{equation}
L_X \approx 1.8 \times 10^{38} \left( {M \over 1.4 M_\odot} \right)
\left( {\dot{M} \over \dot{M}_\mathrm{Edd} } \right) \mbox{  erg/s}
\end{equation}
From these estimates we see that, for accretion at a fraction $\epsilon$ of the 
Eddington rate, eg. $\dot{M} = \epsilon \dot{M}_\mathrm{Edd}$, the magnetic field
must be accounted for (in the sense that $R_M>R$) as long as it is stronger than 
\begin{equation}
B_0 \ge 1.6 \times 10^8 \epsilon^{1/2} \mathrm{ G}
\end{equation}
Since observations indicate that rapidly
rotating neutron stars have magnetic fields of the order of $10^8$~G, 
and many transient LMXBs accrete with $\epsilon \sim 0.01$, 
we infer that the magnetic field is likely to play a role in these systems.

The interaction between a geometrically thin disk and the 
neutron star magnetosphere is a  key ingredient in the standard model 
for accretion. The basic picture is that of a rotating magnetised
neutron star surrounded by a magnetically threaded accretion disk, 
see Figure~\ref{scheme} for a schematic illustration.
In the magnetosphere, accreting matter follows the 
magnetic field lines and gives up angular momentum
on reaching the surface, exerting a spin-up torque.  
The material torque at the inner edge of the disk
is usually approximated by
\begin{equation}
N = \dot{M} \sqrt{GM R_M}
\label{mat1}\end{equation} 
It is important to note that this torque can be significantly 
stronger than the rough estimate for non-magnetic stars.
 
Meanwhile, outside the co-rotation radius,
\begin{equation}
R_c \approx 17 \left( { P \over 1 \mbox{ ms} }\right)^{2/3}  
\left( { M \over 1.4 M_\odot }\right)^{1/3} 
\mbox{ km}
\end{equation}
the field lines rotate faster than the local Keplerian speed, resulting in a negative torque. 
If $R_M>R_c$ the accretion flow will be centrifugally inhibited and matter may be 
ejected from the system. It is easy to see that this will happen 
if the spin period becomes very short, or the rate of flux of material
onto the magnetosphere drops. This is known as the
propeller regime. As accreting matter is flung away
from the star in this phase, the star  
experiences a spin-down torque. To account for this effect we alter the material torque
according to
\begin{equation}
N = \dot{M} R_M^2 [\Omega_\mathrm{K} (R_M) - \Omega] = \dot{M} \sqrt{GM R_M} \left[ 1 - \left({R_M \over R_c} \right)^{3/2}
\right]
\label{mat2}\end{equation}
where $\Omega$ is the spin frequency of the star and
and $\Omega_\mathrm{K}$ is the angular 
velocity of a particle in a Keplerian orbit;
\begin{equation}
\Omega_\mathrm{K}(r) = \left( {GM \over r^3} \right)^{1/2}
\end{equation}
Even though this expression only accounts for the propeller regime in a 
phenomenological way, it agrees with the expectation that
accretion will not spin the star up beyond the point 
 $R_M = R_c$. This leads to the equilibrium period
\begin{equation}
P_\mathrm{eq} \approx 0.30 \left( { B_0 \over 10^{8} \mbox{ G} }\right)^{6/7}
\left( { R \over 10 \mbox{ km} }\right)^{18/7} 
\left( { M \over 1.4 M_\odot }\right)^{-5/7}
\left( { \dot{M}  \over \dot{M}_\mathrm{Edd} }\right)^{-3/7} \mbox{ ms}
\label{Peq1}\end{equation}
Conversely, given an observed spin period we can (assuming that the system is at equilibrium)
deduce the neutron star's magnetic field.

Let us now compare this estimate with the observational data,
summarised in Table \ref{LMXBdata}. To do this we need both stellar
spin rates and an estimate of accretion rates.  Let us first consider
spin rates.  For the X-ray pulsars, we will use the measured pulsar frequency
$\nu_\mathrm{psr}$.  For those sources that are not pulsars but have
burst oscillations, we will assume that the measured burst oscillation
frequency $\nu_\mathrm{burst}$ is the stellar spin frequency.  For the
third class of sources, which exhibit neither pulsations nor
burst oscillations, we will use the separation of the kHz
QPOs, $\Delta \nu_\mathrm{QPO}$, as an
estimate of the stellar 
spin.  As can be seen from those pulsars and burst oscillation
sources that also have kHz QPOs, this estimate is not exact, as the
spin frequency can be as much as double the kHz QPO separation.  In
addition the kHz QPO separation is variable.  For the purposes of this
paper, however, we will estimate stellar spin as the midpoint of the
observed range of $\Delta \nu_\mathrm{QPO}$ for all sources that do
not show burst oscillations or pulsations.  This places an upper limit
on the inferred magnetic field. The inferred field would be lower if the true
spin rate were greater than $\Delta \nu_\mathrm{QPO}$.  

The accretion rate can be estimated from the X-ray luminosity, which
can be highly variable.  It is clear that the estimated equilibrium
period is shortest when the accretion rate is highest (alternatively
for a given spin rate the inferred magnetic field is maximal).  In this paper we
will assume
that the observed spin rate is the equilibrium period 
associated with the maximum accretion rate for a given source, even
for sources that are transient or highly variable\footnote{See
  however \citet{lam04} for a discussion of whether neutron stars do
  reach equilibrium.}. This make sense if one assumes that the main 
contribution to the spin-up torque is associated with the phase when the star 
accretes at the fastest rate.
Hence, the accretion rates given in Table
\ref{LMXBdata} are estimates of maximum accretion rates.  In addition
we will assume $R=10$~km and 
$M=1.4M_\odot$ for all systems. 

\begin{figure}
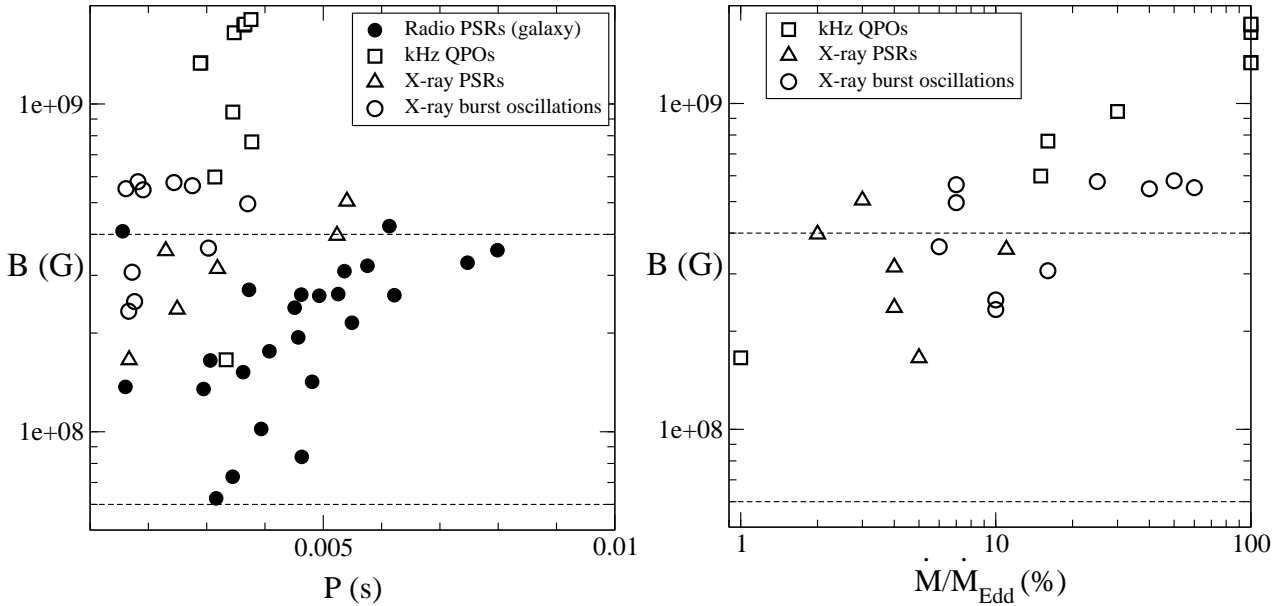

\centerline{
\includegraphics[height=8cm,clip]{bfield_bw_modnew.eps}
\includegraphics[height=8cm,clip]{bfield_mdot.eps}}
\caption{Comparing the neutron stars in LMXBs to the millisecond radio pulsar 
population.  We include 
all millisecond radiopulsars (with periods below 10~ms) in the galaxy. Millisecond pulsar in globular clusters are 
excluded since a significant sample of them are seen to spin up, an effect likely 
due to motion relative to the core of the globular cluster \citep{phinney}. 
which makes the magnetic fields inferred for them dubious. In the left panel we compare the 
inferred magnetic field for the galactic millisecond pulsars, to those 
inferred for accreting neutron stars using the simplest estimate for the 
spin-equilibrium [$B(P_\mathrm{eq})$ is inferred from Eq.~(\ref{Peq1})]. The radio pulsars are shown as filled circles,
systems showing burst oscillations are represented by open circles, data from 
systems where the spin period is estimated from the kHz QPO separation are 
open squares and the accreting X-ray pulsars are shown as open triangles.
We also indicate the (rough) range of magnetic fields for the galactic 
radio pulsars $6\times10^7-4\times10^8$~G. The right panel relates the inferred magnetic fields for the accreting systems
to the accretion rate (in \% of the Eddington rate). This figure indicates that the fields are most seriously
overestimated for the fastest accreting systems. 
[Radio pulsar data taken from the radio pulsar catalogue
http://www.atnf.csiro.au/research/pulsar/psrcat/. Accreting neutron star data determined from Table~\ref{LMXBdata}. ]} 
\label{disk}
\end{figure}

Figure ~\ref{disk} compares the model's predictions for LMXB magnetic
fields with the inferred magnetic fields for the millisecond radio
pulsars.  The agreement is good for LMXBs accreting at the level of $10^{-2}\dot{M}_\mathrm{Edd}$ and below. 
The model does not, however, perform well for systems accreting with $\dot{M}\approx \dot{M}_\mathrm{Edd}$. 
The figure shows that the estimated magnetic fields 
appear to be too large for the systems accreting near the Eddington limit, mainly 
objects for which the spin rate was inferred from the kHz QPO separation.
This is, essentially, the conclusion drawn by \citet{bildsten98}. There seems to be a need
for an additional spin-down torque in the systems that accrete at near-Eddington rates.
Note that, if we had assumed that the 
true spin-rate was twice $\Delta \nu_\mathrm{QPO}$, as indicated in some of the systems 
exhibiting burst oscillations, then the inferred magnetic field would be roughly 
half those indicated, which would still be problematic. 

\begin{table}
\begin{center}
\begin{tabular}{|l|c|c|c|c|c|}
\hline
Source  & Source type & $\nu_\mathrm{psr}$ (Hz) & $\nu_\mathrm{burst}$ (Hz) &
$\Delta \nu_\mathrm{QPO}$ (Hz)  & $\dot{M}/\dot{M}_{Edd}$ (\%)\\
\hline
SAX J1808.4-3658 & P(T) & 401 [1] & 401 [2] & $\sim 200$ [3]  &  4 [4]\\

XTE J1751-305 & P(T) & 435 [5] & & & 11 [4]\\

XTE J0929-314 & P(T) & 185 [6] & & & 3 [4]\\

XTE J1807-294 & P(T) & 191 [7] & & $\sim 190$ [8] & 2 [4]\\

XTE J1814-338 & P(T) & 314 [9] & 314 [10] &  & 4 [4] \\

IGR J00291+5934 & P(T) & 599 [11] & & & 5 [12] \\

\hline

4U 1608-522 & A(T) & & 619 [13] & 225--325 [14] & 60 [15]\\

SAX J1750.8-2980 & A(T) & & 601 [16] & $\approx 317$ [17] & 10 [15] \\

4U 1636-536 & A & & 582 [18] & 242--323 [19]  & 16 [15] \\

MXB 1658-298 & U(T) & & 567 [20] & & 10 [15]  \\

Aql X-1 (1908+005) & A(T) & & 549 [21] & & 50 [15]\\

KS 1731-260 & A(T) & & 524 [22] & 250--270 [23] & 40 [15] \\

SAX J1748.9-2021 & U(T) & & 410 [24] & & 25 [15] \\ 

4U 1728-34 & A & & 363 [25] & 274--350 [26] & 7 [15] \\
 
4U 1702-429 & A & & 330 [27] & 328--338 [27] & 6 [28]\\

4U 1916-053 & A & & 270 [29] & 290,348 [30] & 7 [30,31] \\

\hline
GX 340+0 (1642-455) & Z & & & 280--410 [32] & $\sim 100$ [28]\\

Cyg X-2 (2142+380) & Z & & & 346 [33] & $\sim 100$ [28]\\

4U 1735-44 & A & & & 296--341 [34] &  15 [28] \\

4U 0614+09 & A& & & 240--360 [35] & 1 [28] \\ 

GX 5-1 (1758-250) & Z & & & 232--344 [36] & $\sim 100$[28]  \\

4U 1820-30 & A& & & 230--350 [37] & 30 [28] \\

Sco X-1 (1617-155) & Z & & & 240--310 [38] & $\sim 100$ [28]\\

GX 17+2 (1813-140) & Z & & & 239--308 [39] &  $\sim 100$  [28]\\

XTE J2123-058 & A(T) & & & 255--275 [40,41] & 16 [40,41]\\

GX 349+2 (1702-363) & Z & & & 266 [42] & $\sim 100$ [28]\\

\hline
\end{tabular}
\end{center}
\caption{Data for rapidly rotating neutron stars (with spins
  above 100 Hz), with references
  given in square brackets.  Source type classifications
  are P (pulsar), A (Atoll), Z (Z source) or U (Unknown)
  \citep{has89, van04}.  (T) indicates that the source is transient.    The frequencies given are pulsar spin frequency
  ($\nu_\mathrm{psr}$), burst oscillation frequency
  ($\nu_\mathrm{burst}$) and separation between the two kHz
  Quasi-Periodic Oscillations ($\Delta \nu_\mathrm{QPO}$). The accretion
  rates shown are estimates of maximum accretion rate, as discussed in
  the main text.  References:  [1]
  \citet{wij98}, [2] \citet{cha03}, [3] \citet{wij03}, [4] \citet{gal04},
[5] \citet{mar02}, [6]
  \citet{rem02}, [7] \citet{mar03a}, [8] C.B.Markwardt, private
  communication, [9]
  \citet{mar03b}, [10] \citet{str03}, [11] \citet{mar04}, [12] \citet{gal05}
[13] \citet{har03},[14] \citet{men98}, [15]  D.K.Galloway, private
  communication, [16] \citet{kaa02}, [17] \citet{nat99}, 
[18] \citet{gil02}, [19]  \citet{jon02}, 
[20] \citet{wij01b}, [21] \citet{zha98},[22] \citet{smi97}, [23] \citet{wij97},
 [24]  \citet{kaa03}, [25]  \citet{str96}, [26]
  \citet{mig03},  [27] \citet{mar99}, [28] \citet{for00},
[29] \citet{gal01}, [30]
  \citet{boi00}, [31] \citet{sma88}, [32] \citet{jon00}, [33]
  \citet{wij98b}, [34] \citet{for98}, [35] \citet{van00}, [36]
  \citet{jon02b},[37] \citet{zha98b}, [38] \citet{men00}, [39]
  \citet{hom02},  [40] \citet{hom99}, [41] \citet{tom99}, [42] \citet{zha98c}   }
\label{LMXBdata}
\end{table}

\section{A magnetically threaded disk}

The interaction between an accretion disk and a spinning compact object involves
much poorly known physics. The key issues were discussed in a number of seminal 
papers in the late 1970s (\citet{gho77, gho78, gho79a, gho79b}, see also \citet{frank} for an excellent introduction). 
Although much effort has been invested in this area of research since then --- after all, 
accretion is a cornerstone of astrophysics --- these early papers remain the 
``standard'' description of the problem.  

In this Section we will focus on the contribution to the accretion torque from a
magnetically threaded, thin disk. Our description is based on the work
by \citet{wang87,wang95}
and \citet{ywv97} (see also \citet{yw98} and \citet{yg98}). 

We begin by pointing out that our previous description of the accretion problem
was somewhat inconsistent since our various estimates, eg., of the size of the 
magnetosphere, were based on spherical infall of matter. The model can be improved, 
albeit at the cost of introducing several largely unknown parameters. First of all, 
we need a description of the viscosity in the disk. Viscosity is the main agent that 
dissipates energy and angular momentum, and thus enables matter to flow 
towards the central object. Since the microphysical viscosity (likely 
due to the magnetorotational instability in some form) is difficult to characterise,
it is common to use the so-called $\alpha$-viscosity introduced by 
\citet{shakura}, i.e. let the kinematic viscosity be parametrised as
\begin{equation}
\nu = \alpha c_s  H = \alpha { c_s^2 \over \Omega_\mathrm{K}}
\end{equation}
Here $c_s$ is the sound speed in the disk and  $H\sim c_s/\Omega_\mathrm{K}$ 
is the vertical scale height.
In this description, $\nu$ is a function of $r$ since both $c_s$ and 
$\Omega_\mathrm{K}$ vary with position, but $\alpha$ is taken to 
be constant. In effect, this leads to a model where the
viscosity ensures that the disk remains Keplerian as matter and angular momentum 
is transferred through the disk.

\begin{figure}
\centerline{\includegraphics[height=10cm,clip]{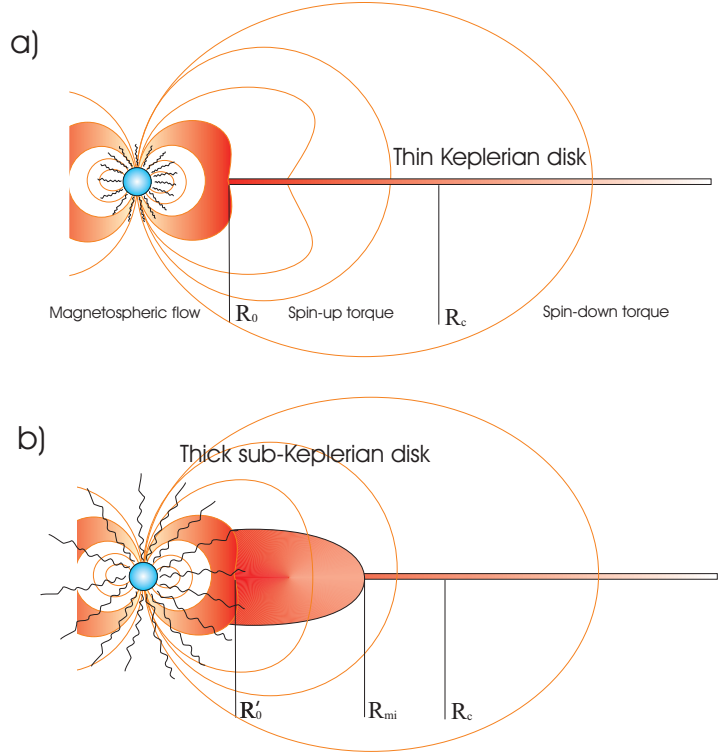} }
\caption{A schematic illustration of the accretion problem for a magnetically 
threaded disk. a) The standard thin disk picture, see \citet{frank}. b) The proposed model for rapidly accreting systems. 
Radiation pressure leads to a thick, sub-Keplerian disk in the inner region  (between $R_{mi}$ and $R_0^\prime$).}
\label{scheme}
\end{figure}

In the case of a magnetically threaded disk, we need to provide a description of 
the interaction between the disk flow and the magnetic field.  Figure~\ref{scheme} 
provides a schematic illustration of the problem. To provide a detailed model of this 
complicated physics problem is, however,
not a simple task. Nevertheless, one may hope that a somewhat simplistic description will be 
able to capture the main features of the complete problem. 

From the 
$\varphi$-component of the Euler equations for the disk flow
we  can estimate the radius at which 
magnetic stresses balance the material 
stresses. We thus find 
\begin{equation}
\dot{M} { d \over dr} \left[\Omega_\mathrm{K} (r) r^2 \right] = - r^2 B_\varphi B_z
\label{phibal}\end{equation}  
where the mass transfer rate $\dot{M}$ will be assumed constant throughout the 
disk. This relation illustrates the difficulty involved in constructing a 
consistent model. If we consider a thin accretion disk, then the $z$-component
of the magnetic field can be taken to be that associated with a 
rotating dipole (with dipole moment $\mu$)
\begin{equation}
B_z = - { \mu \over r^3} = - B_0 \left({R\over r}\right)^3 
\end{equation} 
where $B_0$ is the surface field of the star.
Even though the field may be much more complicated close to the stellar surface
the dipole contribution will dominate far away. The problem is associated with $B_\varphi$. 
This component, which vanishes in the absence of a disk, represents the degree to 
which the magnetic field is dragged along with the matter flow. It is this
interaction which leads to 
the torque on the star that we are aiming to model.

From the MHD induction equation we find that \citep{mestel}
\begin{equation}
\partial_t B_\varphi \approx { B_\varphi \over \tau_\varphi} \approx \nabla \times ( \vec{v} \times \vec{B} ) 
= \gamma ( \Omega - \Omega_\mathrm{K} ) B_z
\end{equation}
where the star (and the magnetic field) is rotating at the constant rate $\Omega$. 
In this equation, it is assumed that the disk flow changes from 
quasi-rigid to Keplerian over a lengthscale $R/\gamma$, with $\gamma \ge 1$. 
\citet{wang95} has considered 
several different mechanisms for the timescale $\tau_\varphi$ (and by implication the 
toroidal component of the magnetic field). He concludes that the various models
lead to quite similar predictions for the accretion torque. This is fortunate, 
since it means that the model is not very sensitive to the unknown 
physics. 
Here we will assume that the main mechanism that prevents the 
field from being dragged along with the flowing matter is turbulent diffusion. 
This leads to \citep{wang95}, 
\begin{equation}
\tau_\varphi \approx {H \over \alpha c_s} \approx { 1 \over \alpha \Omega_\mathrm{K}}
\end{equation}
and consequently
\begin{equation}
B_\varphi \approx { \gamma \over \alpha} { \Omega - \Omega_\mathrm{K} \over \Omega_\mathrm{K} } B_z
\end{equation}

We can now return to Eq.~(\ref{phibal}) and determine the ``inner'' edge of the 
accretion disk $R_0$, at which the matter flow departs significantly from 
a Keplerian profile;
\begin{equation}
\left( { R_0 \over R_c} \right)^{7/2} = { 2 N_c \over \dot{M} \sqrt{G M R_c} } \left[
1 - \left( { R_0 \over R_c} \right)^{3/2} \right] 
\end{equation}
where $R_c$ is the co-rotation radius and we have defined
\begin{equation}
N_c = { \gamma \over \alpha }  { \mu^2 \over R_c^3  } =  { \gamma \over \alpha }  B_z^2 R_c^3 =  { \gamma \over \alpha }  B_0^2 \frac{R^6}{R_c^3}
\end{equation} 
Where $B_z$ is the magnetic field at $R_c$ and $B_0$ is the field at the surface of the star as before.

We can also account for the torque due to the magnetically threaded 
disk outside $R_0$. The corresponding torque follows (essentially) 
from integrating Eq.~(\ref{phibal}) and we get
\begin{equation}
N_\mathrm{disk} = - \int_{R_0}^\infty B_\varphi B_z r^2 dr = - { N_c \over 3 } 
\left[ 2\left( {R_c \over R_0}\right)^{ 3/2} - \left( {R_c \over R_0}\right)^3 \right] 
\end{equation}
As discussed previously, the region $R_0<r<R_c$ contributes a
(positive) spin-up torque, while the region $R_c<r<\infty$ 
provides a (negative) spin-down torque, cf. Figure~\ref{scheme}.

Finally, assuming that the matter gives up all its angular momentum 
(relative to the frame of the star) upon
reaching $R_0$, i.e. that the matter flows along the field lines like ``beads on a wire'' in the
region where the magnetic field dominates the flow\footnote{In reality the problem is expected to be significantly more complicated, with 
the answer depending on the detailed physics in an extended transition region.} , 
we find that the total accretion torque is
\begin{equation}
N =\dot{M} \sqrt{GM R_0}\left[1-\left( {R_c \over R_0}\right)^{3/2} \right] + N_\mathrm{disk} =   { 1 \over 3} \dot{M} \sqrt{GM R_0}
\left[ {7/2 - 7 (R_0/R_c)^{3/2} + 3 (R_0/R_c)^{3}\over 1 - (R_0/R_c)^{3/2} }\right]
\end{equation}
This result shows that the system reaches spin-equilibrium ($N=0$) when
\begin{equation}
\left( {R_0\over R_c} \right)^{3/2} = { 7 - \sqrt{7} \over 6} \longrightarrow R_0 \approx 0.8 R_c
\end{equation} 
This should be compared to the result of \citet{wang95}. The difference arises from the fact that 
Wang uses Eq.~(\ref{mat1}) rather than Eq.~(\ref{mat2}) for the material torque at the inner
edge of the disk (now at $R_0$  instead of $R_M$).

Having added the spin-down torque exerted on the star by the outer 
parts of the disk we find that the system reaches equilibrium slightly 
before $R_0$ reaches $R_c$. 
Nevertheless, the predicted spin-period at equilibrium
\begin{equation}
P_\mathrm{eq} \approx 0.44 \left( { \alpha \over \gamma} \right)^{-3/7} \left( { B_0 \over 10^{8} \mbox{ G} }\right)^{6/7}
\left( { R \over 10 \mbox{ km} }\right)^{18/7} 
\left( { M \over 1.4 M_\odot }\right)^{-5/7}
\left( { \dot{M}  \over \dot{M}_\mathrm{Edd} }\right)^{-3/7} \mbox{ ms}
\label{Peq2}
\end{equation}
does not differ much from the more naive prediction provided by Eq.~(\ref{Peq1}).
Of course, the actual spin period at equilibrium now depends explicitly on the ratio
$\alpha/\gamma$.  Unfortunately, both these parameters are largely unknown. 
In addition, there are many uncertainties (at the level of factors of order unity) in the 
model. 

In order to proceed we note that
the viscosity parameter $\alpha$ is usually assumed to lie in the range 
$0.01-0.3$ \citep{frank}, while $\gamma$ has been assumed to be of order unity \citep{wang87}. 
If we consider values in this range, what does the model imply for the 
magnetic fields of the accreting LMXB neutron stars? From Eq.~(\ref{Peq2}) we 
find that a canonical neutron star will have equilibrium of 3~ms
if 
\begin{equation}
B_0 \approx 9.4\times 10^8 \left( { \alpha \over \gamma} \right)^{1/2} 
\left( { \dot{M}  \over \dot{M}_\mathrm{Edd} }\right)^{1/2} \mbox{ G}
\end{equation} 
We see that for $\alpha/\gamma \approx 0.1$ a star accreting 
at the Eddington rate is predicted to have a magnetic field within the range deduced for the millisecond radiopulsars.
On the other hand, a star accreting at 1\% of this rate will require a larger value of order $\alpha/\gamma \approx 1$ 
in order to lie in the range indicated in Figure~\ref{disk}. 
This means that the inclusion of the torques from a magnetically threaded
thin disk is, in principle, sufficient to remove the direct need for an addition spin-down mechanism 
like gravitational radiation in these systems. Of course, this is achieved at the cost of 
introducing the poorly constrained parameters $\alpha$ and $\gamma$. If we want to
adjust these parameters
in such a way that the inferred magnetic fields agree with those for the radio pulsars
in Figure~\ref{disk}
we essentially need to introduce a suitable $B_\varphi = B_\varphi(\dot{M})$. 
Despite this possibility, we do not think that the thin-disk model is entirely satisfactory. 
As we will argue in the next section, 
additional physics should be included in order to describe the fastest 
accreting systems. In essence, this means that we will only rely on the thin disk 
model for systems accreting  below a few percent of the Eddington rate.
From the above estimates we see that these systems are adequately 
described if we take $\alpha/\gamma \approx 1$. Hence, this will be our canonical 
value from now on.  

\section{Thick disks near Eddington accretion}

The thin disk model we have discussed so far is able to explain many features
of accreting neutron star systems. Yet we will see that it cannot be relied upon 
for rapidly spinning stars accreting near the Eddington limit.
Given this, it is meaningful to ask what the crucial missing piece of physics 
is. At this point, the most naive assumption in our discussion concerns 
the accretion torque arising from the inner edge of the disk, at $R_0$. 
While it seems reasonable to assume that the matter moves along the 
magnetic field lines in the inner region for low rates of accretion, 
it is not so clear that this model will work for faster accretors. 
Several mechanisms may alter the picture. Obvious possibilities 
are: radiation pressure from the emerging X-rays, the near balance between 
centrifugal and gravitational forces for rapidly spinning stars, heating of
the disk in the inner region etcetera. 

\begin{figure}
\centerline{\includegraphics[height=8cm,clip]{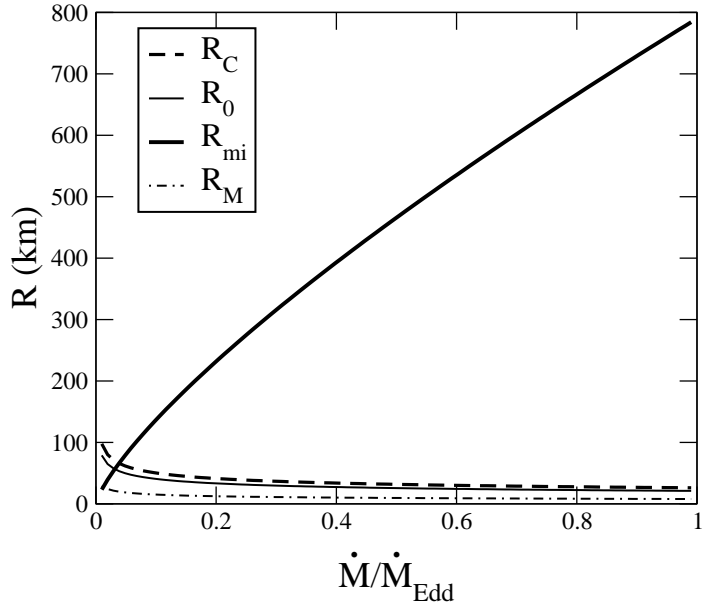} }
\caption{The main lengthscales in the accretion problem. The relevant parameters are taken to be $\alpha/\gamma = 1$, $B_0=10^8$~G
and the star is assumed spin at a rate corresponding to equilibrium for thin disk accretion.
The standard magnetosphere radius (for spherical accretion) $R_M$ is shown as a thin dash-dotted line, and the 
corresponding radius for a thin magnetically threaded disk $R_0$ is a thin solid line. 
The co-rotation radius $R_c$, at which a Keplerian disk co-rotates with the star, is a thick dashed line.
Finally, the distance at which radiation pressure balances gas pressure $R_{mi}$ is shown as a thick solid line.
The figure shows clearly that $R_{mi}>>R_c>R_0>R_M$  above a few
percent of the Eddington accretion rate. This suggests that radiation pressure must be accounted
for, likely leading to a thickening of the disk and a sub-Keplerian flow in the 
inner region. 
} 
\label{Rmifig}
\end{figure}

As a first stab at including these effects we will consider the radiation pressure. 
One can show that radiation pressure balances the gas pressure at a 
radius \citep{frank,paddy}
\begin{equation}
  R_{mi} = 880 \alpha^{2/21} \left( {\dot{M} \over \dot{M}_{\mbox{Edd}} }\right)^{16/21} \left( {M \over M_\odot} 
\right)^{1/3} f^{64/21} \mbox{ km} 
\end{equation} 
where 
\begin{equation}
f= \left[ 1 - \left( { R \over r} \right)^{1/2} \right]^{1/4}
\end{equation}
Strictly speaking, this result holds only for non-magnetic disks, but one can argue that
it remains a good approximation also in the magnetic case \citep{campbell}.
Moreover, it is easy to show that the factor involving $f$ will be near unity 
apart from in the absolute vicinity of the stellar surface. Hence, we 
can use
\begin{equation}
R_{mi} \approx 880 \alpha^{2/21} \left( {\dot{M} \over \dot{M}_\mathrm{Edd} } \right)^{16/21}
\left( {M \over 1.4M_\odot} \right)^{1/3} \mbox{ km}
\end{equation}
as a good approximation. Let us contrast this to the standard radius of the magnetosphere, 
$R_M$. We find that $R_{mi}= R_M$ when
\begin{equation}
\left( {\dot{M} \over \dot{M}_\mathrm{Edd} } \right) \approx 2 \times 10^{-2} \alpha^{-1/11} \left( {B_0 \over 10^8 \mathrm{G} }
\right)^{6/11}
\end{equation} 
The key lengthscales in the problem are illustrated  in Fig.~\ref{Rmifig}. 
This figure shows that, for a neutron star with a weak magnetic field (a typical millisecond pulsar) 
radiation pressure will be important for accretion rates above a few percent of the Eddington rate. 

We thus have to ask how the radiation pressure affects the model outlined in the previous section. 
Phenomenologically, the disk is likely to expand leading to the flow becoming sub-Keplerian. 
In fact, that the thin disk model is unstable in a region where the radiation 
pressure dominates the gas pressure was demonstrated a long time ago by \citet{lightman} (see also 
\citet{shapiro}). 
In order to account for this quantitatively, let us consider the following model. 
The thin disk description is relevant outside $R_{mi}$, and hence describes systems accreting below
the critical rate. For faster accretors, there will exist an inner region inside $R_{mi}$ where 
the disk is no longer thin. To model this region we follow \citet{ywv97}, and assume that 
the flow is such that $\Omega = A \Omega_\mathrm{K}$ with $A\le 1$ \footnote{This model 
should provide an acceptable representation of the inner disk flow, but there are 
important caveats: All radiation dominated configurations tend to be subject to thermal and convective instabilities
and hence may not be stationary, see eg. \citet{szusz} for discussion. }. Before moving on 
we should note that the study of \citet{ywv97} pertains to 
advection dominated accretion \underline{below} a critical accretion rate, while our model concerns 
radiation pressure dominated disks \underline{above} a critical accretion rate. 
This may seem a cause for concern, especially since advection dominated flows are
almost exclusively used in discussions of slowly accreting systems.  
However, as pointed out by \citet{ny94}, the corresponding solution to the equations describing the accretion
problem is likely to be relevant also for rapid accretion. Furthermore, 
the model is sufficiently simple to serve our present purposes. A more detailed analysis
that supports the basic principles behind our model has been carried out by \cite{campbell}.

Repeating the arguments from the thin disk analysis, we find a new co-rotation radius
$A^{2/3} R_c$ and the inner edge of the thick disk region 
$R_0^\prime$ is now determined from
\begin{equation}
\left( { R_0^\prime \over A^{2/3} R_c} \right)^{7/2} = { 2 N_c^\prime \over A^{4/3} \dot{M} \sqrt{G M R_c} } \left[
1 - { 1 \over A } \left( { R_0^\prime \over  R_c} \right)^{3/2} \right] 
\end{equation}
where we have defined
\begin{equation}
N_c^\prime = { \gamma \over \alpha }  { \mu^2 \over A^2 R_c^3  } =  { \gamma \over \alpha }  B_z^2 A^2 R_c^3
\end{equation}
The torque from the inner disk region follows from 
\begin{equation}
N_\mathrm{thick} = - \int_{R_0^\prime}^{R_{mi}} B_\varphi B_z r^2 dr =  { \mu^2 \gamma \over \alpha} 
\int_{R_0^\prime}^{R_{mi}} { 1 \over r^4} \left[ 1 - { 1 \over A }\left({r \over R_c} \right)^{3/2} \right] dr 
\end{equation}
while the outer (thin) disk contributes a torque
\begin{equation}
N_\mathrm{thin} = - \int_{R_{mi}}^\infty B_\varphi B_z r^2 dr =  { \mu^2 \gamma \over \alpha} 
 \int_{R_{mi}}^\infty { 1 \over r^4} \left[ 1 - \left({r \over R_c} \right)^{3/2} \right] dr 
\end{equation}
Working out the algebra, we find that the total torque can be written
\begin{equation}
N = \dot{M} \sqrt{ GM R_0^\prime} { A \over 1 - \bar{\omega} } \left\{  { 7 \over 6} - {7  \bar{\omega}\over 3} + {\bar{\omega}^2}
+ { A (1-A) \over 3} \left({R_c \over R_{mi}} \right)^{3/2} \bar{\omega}^2
\right\} 
\end{equation}
where 
\begin{equation}
\bar{\omega} = { 1 \over A} \left( { R_0^\prime \over R_c}  \right)^{3/2} 
\end{equation}

In this slightly more complicated model, the system will reach spin-equilibrium 
when 
\begin{equation}
 { 7 \over 6} - {7  \bar{\omega}\over 3} + {\bar{\omega}^2}
+ { A (1-A) \over 3} \left({R_c \over R_{mi}} \right)^{3/2} \bar{\omega}^2 = 0
 \end{equation}
Since we must have $R_0^\prime < R_{mi}$ we are always interested in the 
smallest of the two roots to this quadratic. The problem simplifies considerably if
we note that 
\begin{equation}
{ R_c \over R_{mi} } \approx 2 \times10^{-2} \alpha^{-2/21}  \left( {\dot{M} \over \dot{M}_\mathrm{Edd} } \right)^{-16/21}
\left( { P \over 1 \mbox{ ms}} \right)^{2/3}
\end{equation}
for a canonical neutron star, cf. Fig.~\ref{Rmifig}.
 This means that, for a sizeable fraction of the  Eddington accretion rate and 
millisecond spin periods, we have equilibrium when
\begin{equation}
\bar{\omega} \approx { 7 - \sqrt{7} \over 6} 
\longrightarrow R_0^\prime \approx 0.8 A^{2/3} R_c
\end{equation}
From this we can infer the spin period at equilibrium;
\begin{equation}
P_\mathrm{eq} \approx 0.44 A^{-10/7} 
\left( {\alpha \over \gamma}  \right)^{-3/7}  \left( {\dot{M} \over \dot{M}_\mathrm{Edd} } \right)^{-3/7}
\left( {B_0 \over 10^8 \mbox{ G}} \right)^{6/7} \left( { M \over 1.4 M_\odot} \right)^{-5/7}
\left( {R \over 10 \mbox{ km}} \right)^{18/7} \mbox{ ms } 
\label{Peq_final}\end{equation}
This result differs from the thin-disk model only by the factor of $A$. However, it
is easy to see that  this is a key factor which may lead to considerable differences in
the predicted equilibrium spin periods. 



To complete the thick disk model, we need to estimate the coefficient $A$ which 
describes the nature of the sub-Keplerian flow. To do this, we consider the radial 
Euler equation which (for a thin disk) can be approximated by \citep{frank}
\begin{equation}
v_r { \partial v_r \over \partial r} - { v_\varphi^2 \over r}
\approx - { 1 \over \rho} { \partial \over \partial r} \left( p + { B^2 \over 8\pi} \right) 
- { GM \over r^2} + {  B_\varphi^2 \over 4 \pi \rho r}
\label{euler}  
\end{equation}
This equation will remain approximately relevant in the case of a thick disk provided that it is 
interpreted as a height average \citep{ny95a,ny95b}. Apart from very near the Eddington accretion rate 
the dominant velocity component is $v_\varphi$. The situation near $\dot{M}_{\rm Edd}$
is complicated by the fact that the matter in the disk becomes highly virialised.
In our thick disk model, we expect the radiation pressure to dominate in the region $R_0^\prime<r<R_{mi}$.
(It is worth noting the difference between the radial and azimuthal Euler equations here. 
In the latter the axisymmetric radiation pressure will not play a role and the magnetic
and viscous 
stresses dominate.)

We express the radiation pressure gradient in terms of the co-moving radiation flux $L_{co}$ 
\citep{miller90,mitra}
\begin{equation}
{ dp_{\rm rad} \over dr} = - { \kappa \rho \over c} { L_{co} \over 4 \pi r^2} 
\end{equation} 
where $\kappa$ is the opacity of the matter. Since the Eddington luminosity follows from
\begin{equation}
L_{\rm Edd} = { 4\pi G M c \over \kappa} 
\end{equation}
we have
\begin{equation}
{ dp_{\rm rad} \over dr} = - \rho { GM \over r^2} { L_{co} \over L_{\rm Edd}}
\end{equation}
Using this relation in Eq.~(\ref{euler}) we see that the velocity profile 
becomes sub-Keplerian with
\begin{equation}
v_\varphi \approx A \sqrt{ { GM \over r}} \qquad 
\mbox{where} \qquad 
A = \sqrt{ 1 - { L_{co} \over L_{\rm Edd}} } 
\end{equation}
As a rough approximation we can assume that the co-moving flux is equal to the 
stationary flux observed at infinity $L_X = GM \dot{M}/r$ where $r$ is the 
distance to the source. Then 
\begin{equation}
{ L_{co} \over L_{\rm Edd}} \approx { \dot{M} \over \dot{M}_{\rm Edd}}
\end{equation}
and we see that 
\begin{equation}
v_\varphi \approx A r \Omega_K \quad \mbox{ with } \quad
A = \sqrt{ 1 - { \dot{M} \over \dot{M}_{\rm Edd}} } 
\end{equation}
(effects due to eg. beaming have obviously not been included in this estimate).

\begin{figure}
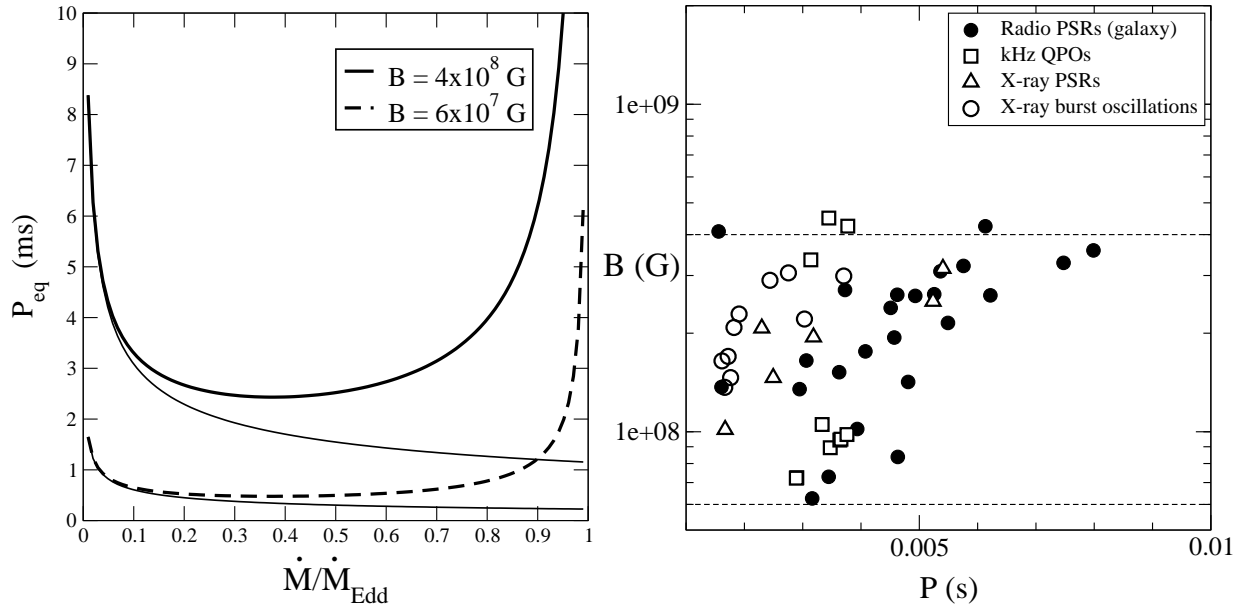

\centerline{\includegraphics[height=8cm,clip]{Peqnew.eps}
\includegraphics[height=8cm,clip]{bfield_thick_modnew.eps} }
\caption{The predicted spin periods at equilibrium for the thick disk model, for 
$\alpha = \gamma = 1$.
In the left panel we show $P_{\rm eq}$ as function of the accretion rate for 
magnetic fields which bracket the range for the  millisecond radio pulsars:
$B= 6\times10^7$~G (thick dashed curve) and $B= 4\times10^8$~G (thick solid curve).
For comparison we also show the prediction of the naive model where 
spinup ceases at $R_c=R_M$ (thin solid curve). The right panel compares the 
inferred magnetic fields for LMXBs to those of the radio pulsars and should be compared
to the right panel in Figure~1.
} 
\label{Peq_fig}
\end{figure}

The results we obtain by 
combining this approximation with the predicted equilibrium period for the 
thick disk model are illustrated  in Figure~\ref{Peq_fig}. 
This figure shows that the thick disk model 
leads to significantly longer equilibrium spins for rapidly accreting systems. 
Conversely, we can use Eq.~(\ref{Peq_final}) to deduce a system accreting 
at 90\% of the Eddington rate and which is observed to spin with a 3~ms period, 
should have a magnetic field of $B\approx 1.4 \times 10^8$~G. A field of this strength
would put this system well within the range of fields inferred for the millisecond radio
pulsars, cf. Figure~\ref{Peq_fig}. The figure shows that our thick disk model leads to 
predicted magnetic fields for the LMXBs which accord well with those of the galactic
millisecond radio pulsars. (In order to infer the magnetic fields shown in Figure~\ref{Peq_fig}
we have assumed that the fastest accreting systems have $\dot{M}= 0.95 \dot{M}_\mathrm{Edd}$. 
This is somewhat ad hoc, but it should be noted that the model breaks down, in the sense that $A\to 0$
which leads to 
$P_\mathrm{eq}$ diverging, as $\dot{M} \to \dot{M}_\mathrm{Edd}$.
There is also significant uncertainty in the accretion rates given
in Table~\ref{LMXBdata}.)

The fact that radiation pressure will affect accretion disk structure,
and hence the spin period of neutron stars in LMXBs, has previously
been discussed by several authors (\citet{whi88, gho91,
  gho92}, see also \citet{mil98, psa99}).  These models, like ours,
give lower inferred magnetic fields for high accretion rate sources
when radiation pressure is taken into account. One issue associated
with the previous models is that if one assumes spin equilibrium, the
models predict a strong
correlation between magnetic field
and accretion rate (see for example Eq.~(25) of \citet{mil98}).
No direct measurements of LMXB magnetic fields have yet been made, so
this correlation cannot be tested, but the physical basis for such a
strong relation is at best unclear. In fact, this has been one of the arguments against
magnetic spin equilibrium models \citep{bildsten98, ush}.  The model
outlined in this paper also predicts a correlation between magnetic
field and accretion rate.  The 
  ``dependence'' of $B$ on $\dot{M}$ is however weaker, due to the
  dependence on accretion rate of the factor $A$.  This illustrates that small
  modifications to the accretion model may be able to remove some of
  the perceived difficulties associated with magnetic equilibrium
  models. 

We conclude this section with a brief discussion of the
observational consequences of this model with regard to the detection
of X-ray pulsars.  Naively one expects
the X-ray pulsars to have higher inferred magnetic fields than the other,
non-pulsing, LMXBs \citep{cum01}.  As is clear from Figure \ref{Peq_fig}, the thick
disk model does not lead to the pulsars 
clustering at higher magnetic fields than the other sources.  One
possibility, suggested by \citet{tit02}, is that
we are prevented from seeing pulsations in many systems due to
atmospheric scattering.  A preliminary study by \citet{kra04} suggests
that the scattering hypothesis may not be borne out by the data, but
this is an area of ongoing research.

\section{Conclusions}

We have discussed the accretion spin-equilibrium for 
neutron stars in LMXBs. The outcome of this study is a more detailed model 
of the accretion torques and an appreciation that it is possible to 
construct a reasonably simple and consistent model for these systems without invoking 
additional spin-down torques due to, for example, gravitational radiation. 
This result is  not particularly surprising. After all, 
the accretion problem is extremely 
complex \citep{frank}, and the torques considered in the studies that argued for the 
need for an additional spin-down mechanism (see \citet{wz97} and  \citet{bildsten98}) were somewhat simplistic. 

Of course, our results should not be taken as proof 
that the LMXBs do not radiate gravitational waves.  
The various proposed mechanisms for generating asymmetries in rapidly
spinning, accreting neutron stars remain (essentially) as viable as before. The key 
difference is that we have eliminated the rationale for locking the gravitational
radiation luminosity to the non-magnetic torque $\dot{M}\sqrt{GMR}$, which has been 
used as an order of magnitude estimate in most studies to date. In our picture, one would not 
be able to infer how the
spin down due to gravitational radiation combines with the accretion torque
from the observed spin periods. This alleviates some ``problems''
with the gravitational-wave models. In the case of accretion induced asymmetries
in the crust \citep{bildsten98}, one can show that the quadrupole deformation required to
balance accretion is
\begin{equation}
\epsilon \approx 10^{-7} \left( {\dot{M} \over \dot{M}_\mathrm{Edd}} \right)^{1/2} 
\left( { P \over 3 \mbox{ ms} } \right)^{5/2}
\end{equation}  
This should be compared to the maximum deformation that the crust can sustain, 
which according to \citet{ush} can be approximated as
\begin{equation}
\epsilon_\mathrm{max} < 5 \times 10^{-7} \left( { u_\mathrm{break} \over 10^{-2} } \right) 
\end{equation}
where the breaking strain $u_\mathrm{break}$ is usually (based on results for terrestrial materials)
assumed to be in the range $10^{-4} - 10^{-2}$. These estimates show that the breaking strain must be
near the upper limit of the expected range in order for these asymmetries to balance near Eddington 
accretion in a star spinning at a period of a few milliseconds. By weakening the accretion torque, 
while at the same time not 
altering the mechanism generating the asymmetry (eg. the accretion rate),  
this issue is made less critical. 

Our results also impacts the suggestion that the gravitational waves are emitted 
by unstable r-mode oscillations in the stellar fluid. In this case, 
the r-modes are expected to become unstable below a critical rotation period $P_\mathrm{crit}$.
The point at which the instability becomes relevant depends on many complicated issues
concerning viscosity, superfluidity etcetera (see \citet{narev} for a discussion)
but it is plausible that $P_\mathrm{crit} \approx 2-3$~ms. In the context of the present model, we obviously 
need $P_\mathrm{crit}>P_\mathrm{eq}$ in order for the r-mode instability to be relevant. 
Considering the results illustrated in the left panel of Figure~\ref{Peq_fig}, 
we expect that the instability may come into operation in weak magnetic field systems which 
are neither very slow nor very fast accretors. 

The most important next step in modelling the LMXBs concerns the variability 
in the spin with varying accretion rate. The spin of accreting X-ray pulsars is known to 
vary considerably \citep{bildsten97} on a timescale which is roughly similar to the 
variations in the accretion rate. But the data also suggests that there may not be a direct link
between increased X-ray flux and an increase in the spin-up torque. It is important to understand this
variability in general. This is also a very important issue for attempts to search for gravitational
waves from the LMXBs. Any variability on timescales shorter than the observation time that remains 
unaccounted for will likely lead to a significant loss in signal-to-noise ratio. Our aim is to 
turn our attention to this challenging problem in the near future.

\section*{Acknowledgements}

We thank K.D. Kokkotas, D. Chakrabarty and J.C. Miller for useful discussions.  
We are especially grateful to D.K.Galloway for valuable
advice and for providing us with LMXB
accretion rates.
NA also thanks T.E. Strohmayer and NASA GSFC for generous hospitality, 
and interesting discussions about LMXBs.
This work was supported by PPARC grant PPA/G/S/2002/00038.  ALW holds
a National Research Council Resident Research Associateship at NASA
GSFC.  


\end{document}